\documentclass[journal,twoside,web]{ieeecolor}
\usepackage{generic}
\usepackage{cite}
\usepackage{amsmath,amssymb,amsfonts}
\usepackage{algorithmic}
\usepackage{graphicx}
\usepackage{textcomp}
\def\BibTeX{{\rm B\kern-.05em{\sc i\kern-.025em b}\kern-.08em
    T\kern-.1667em\lower.7ex\hbox{E}\kern-.125emX}}
\begin{document}
\title{Benchmarking Inverse Rashba-Edelstein Magnetoelectric Devices for Neuromorphic Computing}
\author{Andrew W. Stephan, Jiaxi Hu, and Steven J. Koester, \IEEEmembership{Fellow, IEEE}
\thanks{Manuscript submitted \today. This work was supported in part by C-SPIN, one of the six centers of STARnet, a Semiconductor Research Corporation program, sponsored by MARCO and DARPA.}
\thanks{The authors acknowledge the Minnesota Supercomputing Institute (MSI) at the University of Minnesota for providing resources that contributed to the research results reported within this paper. URL: http://www.msi.umn.edu}
\thanks{The authors wish to thank Dr. Ramamoorthy Ramesh of the University of California, Berkeley, for helpful discussions.}
\thanks{A. W. Stephan is with the College of Science and Engineering, University of Minnesota, Minneapolis, MN 55455 USA (e-mail:steph506@umn.edu).}
\thanks{J. Hu is with the College of Science and Engineering, University of Minnesota, Minneapolis, MN 55455 USA (e-mail:huxxx499@umn.edu).}
\thanks{S. J. Koester is with the College of Science and Engineering, University of Minnesota, Minneapolis, MN 55455 USA (e-mail:skoester@umn.edu).}}

\maketitle

\begin{abstract}We propose a new design for a cellular neural network with spintronic neurons and CMOS-based synapses. Harnessing the magnetoelectric and inverse Rashba-Edelstein effects allows natural emulation of the behavior of an ideal cellular network. This combination of effects offers an increase in speed and efficiency over other spintronic neural networks. A rigorous performance analysis via simulation is provided.
\end{abstract}


\section{Introduction}
\label{sec:introduction}
As complementary metal oxide semiconductor (CMOS) transistors approach their scaling limit\cite{ScalingLimit}, the search for a new technology to fuel next-generation computing has accelerated. Initial attempts to create spintronic logic devices have generally fallen short of the efficiency and performance of CMOS\cite{BeyondBenchmark}. Many designs have been proposed, varying from all-spin logic (ASL) driven by spin-torque\cite{ASL} to hybrid spin and charge devices utilizing various combinations of the magnetoelectric (ME) effect and spin-orbit effects\cite{COMET,MESO1,MESO2,MESO3,LIF}. In particular, the pairing of ME\cite{ME1,ME2,MERef} and Inverse Rashba-Edelstein\cite{IR1,IR2} (IR) effects shows promise for spintronic computing. We herein describe our design for an IR-ME neuron (IRMEN) for non-Boolean neuromorphic computing.

The enormous parallel processing power of cellular neural networks (CNNs) makes them extremely useful for some purposes for which Boolean logic is ill-suited, such as filtering and various recognition tasks\cite{AssociativeMemory,AudioRecognition}. It is thus fitting to apply beyond-CMOS structures to this application. Using the IR and ME effects to interface with a spintronic neuron allows inter-device communication to occur within the charge domain. This opens the door to a powerful pairing of spintronic neurons, with their natural fit to the task of neuromorphic computing\cite{Naeemi1,Naeemi2}, and efficient charge-based synapses in the form of CMOS interconnects. Our devices have an energy barrier sufficient to be nonvolatile. Simulated IRMEN networks performing low-pass filtering indicate they are capable of fast, efficient processing, performing state transitions in tens of picoseconds and network operations with a per-neuron energy-delay product on the order of $10^{-26}$ J-s. The entire CNN also operates asynchronously, avoiding the need for a complex clock distribution network. 

The remainder of this paper is organized as follows. In Section II we give a brief background on CNNs and the physics of our devices. Section III contains a description of the proposed device structure, the simulation process and the parameters used. The state equations governing each neuron are presented as well.  In section IV the simulated IRMEN performance is analyzed in the context of individual neurons and full CNNs. After a discussion of thermal stability and ferroelectric properties as they affect the simulation, possible effects of scaling, and comparison to other spintronic neuromorphic devices in Section V, we conclude in section VI.

\section{Background}
\label{sec:Background}
\subsection{Cellular Neural Networks}
CNNs are a type of neural network that specializes in computing on datasets in which the geometric proximity of data points is highly relevant\cite{CNNs}. They can be thought of as purely analog hybrids of feedforward neural networks and the Ising computing scheme\cite{Ising}. A CNN comprises a grid of regularly spaced neurons, each of which constantly communicates bidirectionally within a local group, or neighborhood. The connections carrying this communication are called synapses, and generally allow for variable weighting of the signals they transmit. The weights applied to a network specify its behavior in response to initial stimuli, defining what function it will perform.

The foundational theory of CNNs was put forward by Chua and Yang\cite{CNNs}.  In their model each neuron state is encoded in the voltage across a capacitor. Synapses consist of voltage-controlled ideal current sources applied to the capacitor and a resistor in parallel. Thus the state each neuron is attempting to reach at any given time is determined by a linear combination of the outputs and sometimes input biases of all neurons within its neighborhood at that moment, accounting for propagation delays. The state equation for the neuron located in row $i$ and column $j$ with state $V_{i,j}$ is 
\begin{gather*}
C \frac{dV_{i,j}(t)}{dt} = -\frac{1}{R}V_{i,j}(t)\\
+ \sum_{k,l} A(i,j,k,l)f(V_{k,l}) \\
\end{gather*}
\vspace{-.5in}
\begin{gather}
+ \sum_{k,l}B(i,j,k,l)U_{k,l} + I_{i,j} 
\label{eq:CNNTheory}
\end{gather}
where the sums are taken over the neighborhood of the neuron $i,j$. The terms $A$, $B$, and $I$ are the weights between the output $f(V_{k,l})$ and its input to $V_{i,j}$, the weights between the net input to $V_{k,l}$ and its input to $V_{i,j}$ and a constant bias current which sets the default state respectively. The output of a neuron is set by a simple transfer function which maps a voltage value $-V_{max} \le V_{i,j} \le V_{max}$ to itself and saturates to $\pm V_{max}$ for values $|V_{i,j}| \ge V_{max}$. Here $V_{max}$ is the maximum possible potential for input and output. We propose an efficient implementation of this design with spintronic neurons and simple CMOS synapses which will perform the functions of a CNN.

\subsection{Magnetoelectric Effect and Interlayer Coupling}
The ME effect has been shown to allow electic field-driven switching of ferromagnetic layers adjacent to multiferroic-antiferromagnet (M-AFM) layers such as bismuth ferrite (BFO). A weak ferromagnetic moment arises in BFO due to the Dzyaloshinskii-Moriya interaction. An electric field applied to the BFO reverses its ferroelectric polarization and, via powerful ME coupling, its ferromagnetic moment as well. The resultant effective magnetic field can cause magnetization reversal of the adjacent ferromagnet (FM) \cite{ME1,ME2}. The precise nature and strength of this interlayer coupling is not fully known. Many device simulations model the coupling as identical to the magnetoelectric coupling between the electric and magnetic polarization of the BFO itself\cite{COMET,MESO1}. Others argue the coupling is mediated by strain or magnetoelastic energy transfer\cite{ME2,Ramesh}. Experiments have yet to show which approach is most accurate. With the mechanism of coupling not fully understood, it is reasonable to model the effect as a simple energy transfer between the polarization energy of the BFO and the FM layer, 
\begin{gather}
\boldsymbol{H}_{ME} = \zeta \frac{2 P_{ME} V_{ME}}{M_s} \boldsymbol{\hat{y}},
\label{eq:ME}
\end{gather}
where $\boldsymbol{H_{ME}}$ is the effective magnetic field, $P_{ME}$ is the total BFO charge polarization, $V_{ME}$ is the capacitor potential, $M_s$ is the magnetic saturation and $\zeta$ is a net efficiency coefficient. Bold font is used to indicate vector quantities. The polarization $P_{ME}$ may have significant hysteresis if the BFO forms a nonlinear capacitor. We assume a simple linear capacitor with no ferroelectric hysteresis. It has been shown that the hysteretic characteristics of BFO can be made close to linear by doping it with barium \cite{BFO1}. In this case the field in steady state is
\begin{gather}
\boldsymbol{H}_{ME} = \zeta \frac{2 C_{ME} V_{ME}^2}{M_s} \boldsymbol{\hat{y}}.
\label{eq:linearME}
\end{gather}

\subsection{Inverse Spin-Orbit Effects}
The IR effect provides a spin-to-charge transduction mechanism to complement the charge-to-spin transformation of the ME effect. When spin-polarized current flows through an appropriate bilayer spin-orbit material interface such as Ag/Bi, some of the spins are redirected in a direction orthogonal to the spin and current flow axes. In an open circuit, this results in a lateral polarization-dependent potential between the ends of the interface which does not significantly affect the potential on the overlying layer providing the current \cite{IR1,IR2,IR3}. The effect can reasonably be assumed to be capable of charging a capacitor up to the IR potential assuming the $CV$ product is not too large. The IR stack is modeled as a dependent voltage source $V_{IR}$
\begin{gather}
V_{IR} = I_d \cdot R_X
\label{eq:VIR}
\end{gather}
with output determined by the incoming drive current $I_d$ and the effective internal resistance
\begin{gather}
R_X= \eta \frac{\lambda}{w_{IR}} (\boldsymbol{M}\cdot\boldsymbol{\hat{y}})R_{IR}^x,
\label{eq:RX}
\end{gather}

where $\lambda$ is the effective conversion length determined by material properties, $w$ is the width of the $IR$ interface along the y-axis, $\boldsymbol{M}$ is the FM magnetization, $R_{IR}^x$ is the total resistance of the IR stack in the x direction \cite{IR1}. The effective resistance term also accounts for the polarization of the drive current due to the FM by including the spin injection efficiency $\eta$. Here we will also define the resistance of the IR stack in the z direction $R_{IR}^z$. Recent research in spin-orbit coupling materials offers hope that spin-orbit materials with high resistivities, and their benefits to spintronic CNNs, may be available soon\cite{HighResistivity}. We note that the bulk inverse Spin-Hall effect or a topological insulator could also be used and, to first order, would produce similar results \cite{ISHE,TIs}. The precise version of inverse spin-orbit effect is not critical. 

\section{Simulation}
\label{sec:Simulation}
\subsection{Neuron Structure}

\begin{figure}
\centering
\includegraphics[scale=0.3]{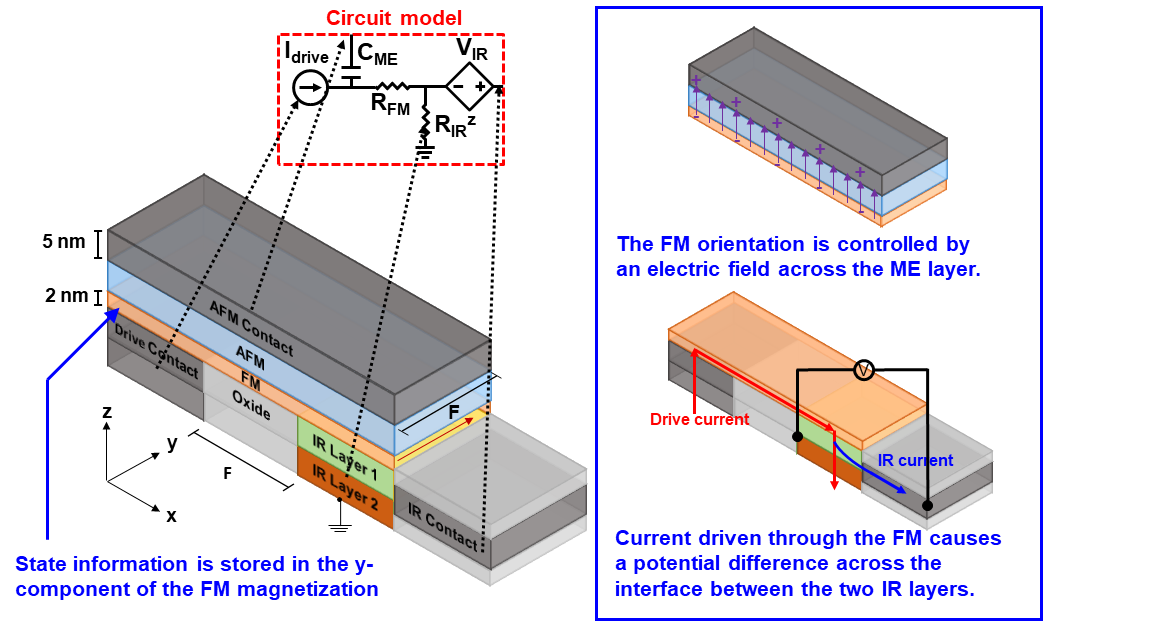}
\caption{Scale diagram of the proposed IRMEN. A nano-ferromagnet encodes the state of the neuron in the y-projection of its magnetization. This magnetic orientation is influenced by the M-AFM in the magnetoelectric capacitor to which it is coupled. Charge current injected into the FM becomes spin-polarized and, via the inverse Rashba-Edelstein effect, provides the voltage for readout.}
\label{fig:IRMENDiagram}
\end{figure}

The IRMEN has a functional similarity to the neurons posited by Chua and Yang\cite{CNNs}. Its structure, inspired by the magnetoelectric spin-orbit (MESO) digital logic device\cite{MESO1,MESO2}, consists of a thin layer of a metallic FM comprising one plate of a capacitor with an M-AFM such as BFO for the spacer. The FM is grounded through an IR bilayer interface. The neuron state, encoded in the y-axis projection of the FM magnetization rather than the capacitor potential, can be written by applying charge to the nonmagnetic electrode of the capacitor which produces an electric field and manipulates the BFO polarization. The magnetic state encoding method has multiple benefits, including lending the neuron nonvolatility and causing it to naturally compute a nonlinear transfer function between input and output. The Stoner-Wohlfarth model predicts a spectrum of possible transfer functions between the saturated linear function used in \cite{CNNs} and a simple step function depending on the angle between the dominant anisotropy axis of the magnet and the direction of applied field. We opted for the latter function type to simplify the simulation. More complex CNN functions requiring sigmoidal transfer functions can be implemented simply by changing the angle between the applied field and anisotropy axes.

\subsection{Synapse Structure}
The neuron state is read by driving current through the FM. This current becomes polarized depending on the orientation $\textbf{M}$ and spin injection efficiency $\eta$ of the FM and produces a magnetization-dependent potential $V_{IR}$ between the x-axis faces of the IR junction (Fig. \ref{fig:IRMENDiagram}). Since the state is constantly being read, it is referenced to the DC potential caused by this drive current and is thus immune to being influenced by it. The IR potential is used to gate a set of CMOS repeaters that act as the synapses of the CNN with outputs joined to the nonmagnetic plates of the capacitors of neighboring neurons. These charge-based synapses prevent signal decay and provide spin isolation between neurons.  A notional 2D IRMEN fabrication layout including the neuron and five output synapses was designed. Though not fully optimized, this layout indicated that the neuron and up to five synapses for the simplest uniform-weight nearest-neighbor network can fit within a compact $14 \times 20$ $F^2$ cell where $F$ is the minimum feature size. Cells in a network requiring non-uniform weights may require nanoresistors or analog memristors with dimensions of 10-100 nm, increasing the area requirement \cite{Memristors,Nanoresistors}.

\subsection{State Equations}
The simulation process treats the FM as a single-domain magnetic vector with net moment $\boldsymbol{M}$ governed by the Landau-Liftshitz-Gilbert (LLG) equation. This gives the IRMEN state equation, analogous to the classic CNN state equation (\ref{eq:CNNTheory}) insofar as it governs the state $\boldsymbol{M_{i,j}}$ of the neuron with coordinates $i,j$ according to its inputs $\boldsymbol{H}_{Ext}$:
\begin{gather*}
\frac{d\boldsymbol{M}_{i,j}}{dt} = \gamma \cdot (\boldsymbol{M}_{i,j}\times (\boldsymbol{H}_I + \boldsymbol{H}_{ME})
\end{gather*}
\vspace{-.2in}
\begin{gather}
 - \alpha \cdot (\boldsymbol{M}_{i,j} \times \boldsymbol{M}_{i,j} \times (\boldsymbol{H}_I + \boldsymbol{H}_{ME}))
 \label{eq:dMdt}
\end{gather}
In this equation $\alpha$, $\gamma$ are the damping constant and gyromagnetic factor, respectively. The $\boldsymbol{H}_I$ term represents the net internal field imposed upon the magnetic moment. The external term is a coupling field proportional to the polarization $P_{i,j}$ of the adjacent BFO layer. This polarization obeys the Landau-Khalatnikov (LK) equation. Preliminary simulations with significant hysteresis and large shifts in total polarization showed switching to within less than $1\%$ of steady-state with a characteristic time of roughly 7 ps. Given the subsequent assumption of linear BFO capacitors and the fact that even significant polarization switching occurs 1-2 orders of magnitude faster than that of the FM, we simplified the simulation by replacing the ferroelectric dynamics with a simple exponential decay equation with a time constant of 7 ps. This significantly increased the allowed time step size, reducing simulation time. 
The internal field term $\boldsymbol{H}_I$ from (\ref{eq:dMdt}) is a sum of the various fields intrinsic to all magnetic materials. This includes the crystalline anisotropy field $\boldsymbol{H_K}$, the random thermal field $\boldsymbol{H_T}$, and the demagnetization field $\boldsymbol{H_D}$. 
\begin{gather*}
\boldsymbol{H_I} = \boldsymbol{H_K} + \boldsymbol{H_T} + \boldsymbol{H_D}
\end{gather*}
The anisotropy field is given by
\begin{gather} 
\boldsymbol{H}_K = \frac{2K}{M_S}  \boldsymbol{\hat{m}}\cdot \boldsymbol{\hat{y}},
\label{eq:HK}
\end{gather}
where $\boldsymbol{\hat{m}}$ is the normalized magnetic moment. The thermal field is given by the typical multivariate Gaussian random variable used in finite-time numeric simulations\cite{Randy}. This variable has zero mean and variance matrix
\begin{gather}
\boldsymbol{\Sigma} = \boldsymbol{I} * \sqrt{\frac{2 k_B T \alpha}{\gamma M_S V \Delta t}},
\label{eq:HT}
\end{gather}
where $\boldsymbol{I}$ is the three-component identity matrix and $k_B$, $T$, $V$, and $\Delta t$ are the Boltzmann constant, temperature, magnetic volume and simulation time step respectively. Finally, the demagnetization field is estimated using the approximation
\begin{gather}
\boldsymbol{H}_D = -   \frac{\{l^{-2}M_x, w^{-2}M_y, t^{-2}M_z\}}{l^{-2} + w^{-2} + t^{-2}},
\label{eq:HD}
\end{gather} where $M_x = \boldsymbol{M}\cdot \boldsymbol{\hat{x}},$ etc. The terms $l$, $w$, and $t$ are the dimensions of the magnet in the $x$, $y$ and $z$ directions respectively. The term $\boldsymbol{H}_{ext}$ in (\ref{eq:dMdt}) represents the contribution to the magnetic vector motion provided by the total input potential, which, in steady state, is approximately a linear combination of neighboring gate potentials. This mimics the sum term $\sum_{k,l}A(i,j,k,l)f(V_{k,l})$ from (\ref{eq:CNNTheory}).

\begin{figure}
\centering
\includegraphics[scale=0.5]{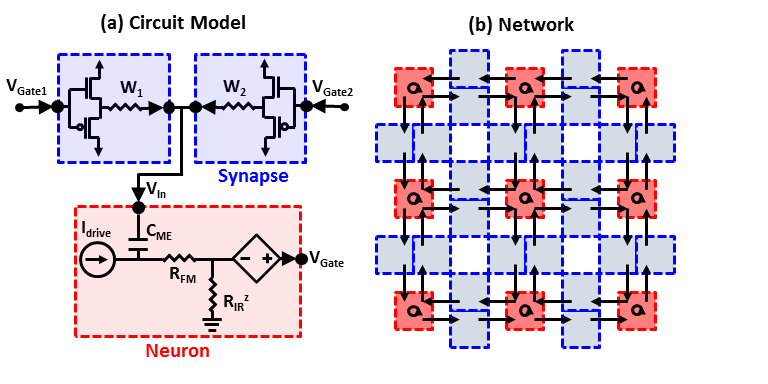}
\caption{ Circuit and network diagrams. (a) Neuron and synapse circuit models including resistive weights. Inputs undergo a passive weighted averaging. (b) Network diagram indicating two-way nearest-neighbor neuron communications via synapses. Each neuron also has a self-influencing synapse indicated by the loop arrows.}
\label{fig:weights}
\end{figure}

The ME capacitor of each neuron is controlled by several parallel synapses depending on the degree of connectivity in the network. These synapses set the capacitor voltage $V_{i,j}$ to a value between the two repeater supply rails based on a nearly linear combination of all states within its neighborhood. We represent the net current provided at the output of a synapse with gate and output potentials $Y_{k,l}, V_{i,j}$ as $g(Y_{k,l},V_{i,j})$. The form of this function was determined by HSPICE simulations using the Arizona State University Predictive Technology Model at the 16nm node \cite{PTM}. The $g$ function also contains the weighting factors, if any, analogous to $A(i,j,k,l)$ from (\ref{eq:CNNTheory}). These weights can be applied via resistors between the CMOS repeater outputs and the target neurons as in Fig. \ref{fig:weights}. As this is a preliminary work to establish the viability of IRMEN CNNs, we did not consider weighting components in the simulation since the data smoothing function does not require nonuniform weights, removing the necessity of nanoresistors in the IRMEN cell. 
For neuron $i,j$ the magnetoelectric potential $V_{i,j}$ is described by
\begin{gather}
\frac{dV_{i,j}}{dt} = \frac{1}{C_{V}}\sum_{k,l}g(Y_{k,l},V_{i,j})
\label{eq:dVdt}
\end{gather}
and the gate potential $Y_{i,j}$ by
\begin{gather*}
\frac{dY_{i,j}}{dt} = \frac{1}{C_{Y}}(\frac{R_Y(\boldsymbol{M}_{i,j})}{R}(I_d+\sum_{k,l}g(Y_{k,l},V_{i,j}))
\end{gather*}
\vspace{-.2in}
\begin{gather}
-\frac{Y_{i,j}}{R_Y(\boldsymbol{M}_{i,j})}),
\label{eq:dYdt}
\end{gather}
where the terms $C_Y$ and $C_V$ refer to the net synapse gate capacitance and magnetoelectric capacitance, respectively.

We used Matlab to numerically solve (\ref{eq:dMdt}-\ref{eq:dYdt}) using the fourth-order Runge-Kutta method. After validating with HSPICE simulations we made the approximation that the steady-state input potential $V_{i,j}$ is proportional to the sum of the gate potentials $Y_{i,j}$ on the input synapses. This leads to 
\begin{gather}
\frac{dV_{i,j}}{dt} \approx \frac{1}{C_{V}R_{V}} \big(\sum_{k,l}(\frac{0.65}{N}Y_{k,l})-V_{i,j}\big),
\label{eq:lineardVdt}
\end{gather}
where $N$ is the size of the neighborhood and $R_V$ is the resistance of the charging circuit due mostly to the repeater FET resistances. These resistances and the proportionality constant $0.65$ were determined empirically by HSPICE. This quantity depends on the applied gate potentials.
All non-fundamental parameters used in the simulation are given in Table \ref{tab:parameters}. The saturation, anisotropy and spin injection efficiency of the FM layers were chosen to be consistent with Heusler alloys \cite{Heusler1, Heusler2, Heusler3}. The conversion length chosen is between the value reported for Ag/Bi interfaces \cite{IR1} and those of novel 2D materials such as MoS2 monolayers \cite{IR3}. The interlayer coupling efficiency $\zeta$ does not yet have experimental verification. Instead we chose a number that provided similar behavior to other simulated ME/FM heterostructures such as ref. \cite{MESO1} and represented an optimistic value that would provide an upper bound on the effectiveness of ME devices.  Energy consumption in the network was tracked throughout the simulation by summing the $I\cdot V$ power of the neuron drive and transient capacitor currents in the drive stack, the rail-to-rail leakage in the synapses, and all magnetoelectric and gate capacitor charging energy. Note that this does not produce the same energy-delay switching product as might be used to benchmark a Boolean logic device since all neurons are supplied with power until the entire network is considered to have completed its computation. Many neurons continue consuming power long after they have finished reached a steady state. When used as the energy-delay switching characteristic for individual devices, therefore, this data provides an overestimate.

\begin{table}
\centering
\caption{Simulation Parameters}
\label{table}
\setlength{\tabcolsep}{3pt}
\begin{tabular}{|p{25pt}|p{80pt}|p{50pt}|}
\hline
Symbol& 
Quantity& 
Value \\
\hline
\vspace{0.005in}
$K$& 
\vspace{0.005in}
crystalline anisotropy& 
\vspace{0.005in}
$6 \cdot 10^5$ erg/cm$^3$\cite{Heusler1, Heusler2, Heusler3} \\
$V_{FM}$& 
ferromagnet volume& 
1536 nm$^3$ \\
$\eta$& 
spin injection efficiency& 
0.9\cite{Heusler1, Heusler2, Heusler3} \\
$M_S$& 
saturation magnetization& 
500 emu/cm$^3$\cite{Heusler1, Heusler2, Heusler3} \\
$C_{ME}$& 
ME capacitance& 
0.68 fF \\
$\zeta$& 
interlayer coupling & 
0.5 \\
$\lambda$& 
spin conversion length& 
1 nm\cite{IR1, IR3} \\
$\rho_{IR}$& 
IR material resistivity& 
1-20 m$\Omega$ $\cdot$ cm\cite{HighResistivity} \\
$\Delta t$& 
time step& 
0.5 ps \\
$V_{drive}$& 
neuron drive voltage& 
300-1000 mV \\
$V_{DD}$& 
synapse supply voltage & 
$\pm$ 500 mV \\
$F$& 
minimum feature size& 
16 nm \\
\hline
\multicolumn{3}{p{200pt}}{Non-fundamental physical parameters used in the simulation. Magnetic units given in CGS.}\\
\end{tabular}
\label{tab:parameters}
\end{table}

\section{Results}
\label{sec:Results}

\begin{figure}
\centering
\includegraphics[scale=0.45]{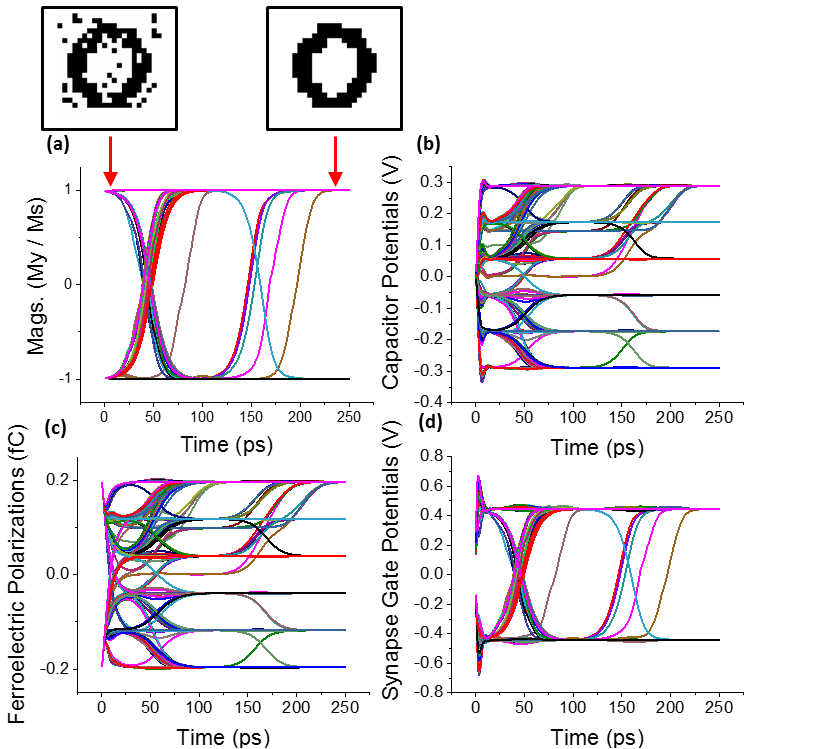}
\caption{ Simulation of an IRMEN CNN filtering noise out of a 21 x 21 pixel image of a circle. This simulation was performed with 1 V drive potential and an IR resistivity of 10 m$\Omega \cdot$cm. (a) Normalized y-projection of every magnetic state vector vs. time. Before and after images are shown above. (b) ME capacitor voltages vs. time. (c) Ferroelectric polarization charges vs. time. (d) IR gate potentials vs. time.}
\label{fig:BigFigure}
\end{figure}

One of the simplest CNN tasks is low-pass filtering, used to clean up noisy images. This is achieved by imposing a simple set of connections on the network in which each neuron receives equally weighted input from itself and its four geometrically nearest neighbors and zero input from all others. Thus in steady state,
\begin{gather}
V_{i,j} = \frac{V_{Max}}{5}(\boldsymbol{m}_{i,j}+\boldsymbol{m}_{i-1,j}+\boldsymbol{m}_{i+1,j}
+\boldsymbol{m}_{i,j-1}+\boldsymbol{m}_{i,j+1})
\end{gather}
where the net synapse output is pinned between $\pm V_{Max}$ as determined by the supply. This is analogous to setting the weights in (\ref{eq:CNNTheory}) such that $A(i,j,i \pm 1,j) = A(i,j,i,j \pm 1) = 1$ while all other weights $A(i,j,k,l) = 0$. In the case of the IRMEN network this is achieved by providing every neuron with identical synapses containing no additional resistances beyond the intrinsic FET resistance. 

An example simulation of an IRMEN CNN filtering noise out of a $21$ $\times$ $21$ binary pixel image of a circle is given in Fig. \ref{fig:BigFigure}. The initial angles of the magnetizations follow the Fokker-Plank distribution with the preferred angle of a given magnet corresponding to its pixel value in the image. We attribute the relatively small magnitude of the thermal fluctuations throughout the process to the fact that $H_{ME} \gg H_T$ with the chosen parameters. The average error percentage is determined by
\begin{gather}
E = \frac{100}{N}\sum_{i,j}|p_{i,j} - \boldsymbol{M}_{i,j} \cdot \boldsymbol{\hat{y}}|,
\end{gather}
where $N$ is the number of pixels and $p_{i,j}$ is the desired value of pixel $i,j$ scaled to $\pm 1$.

\begin{figure}
\centering
\includegraphics[scale=0.44]{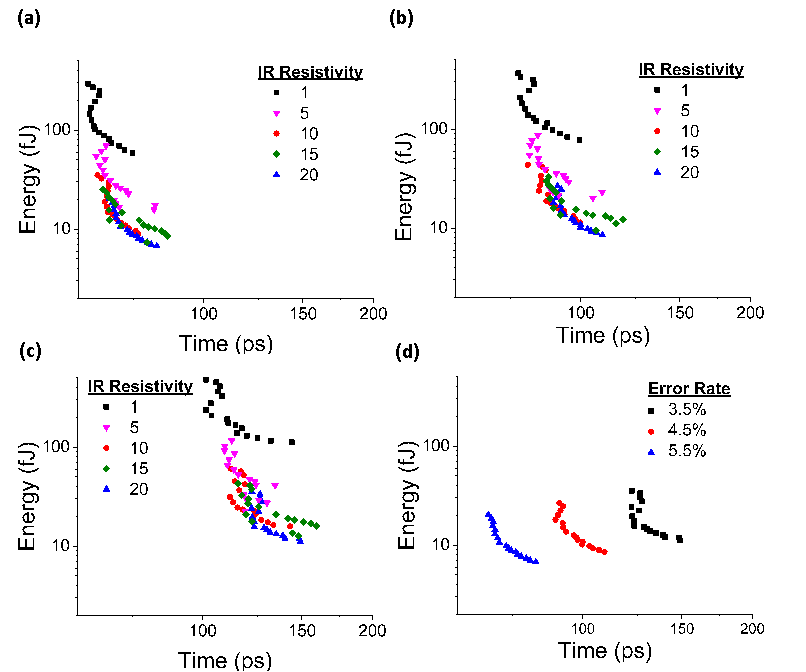}
\caption{ Average energy per cell vs. network delay with IR resistivity in m$\Omega \cdot$ cm and target error percentage as parameters. (a) 5.5$\%$ error. (b) 4.5$\%$ error. (c) 3.5$\%$ error. (d) All target error percentages at 20 m$\Omega$ $\cdot$ cm resistivity.}
\label{fig:EnergyDelayData}
\end{figure}

We now report the relationships between the operational delay, energy cost and various parameters of the CNN when performing this filtering function. With the original clean circle image as reproduced in Fig. \ref{fig:BigFigure}, we simulated filtering with many different drive voltages up to 1 V. Each data point is the result of a 500-iteration Monte Carlo simulation with randomized $10\%$ noise patterns. The average energy consumed by neurons and synapses per cell versus the time required for the network to achieve the target error is plotted in Fig. \ref{fig:EnergyDelayData}. Both target error percentage and spin-orbit material resistivity are used as parameters. This illustrates the tradeoff that exists between energy, time and functional accuracy while showing the pure benefits of increased resistivity. 

\begin{figure}
\centering
\includegraphics[scale=0.48]{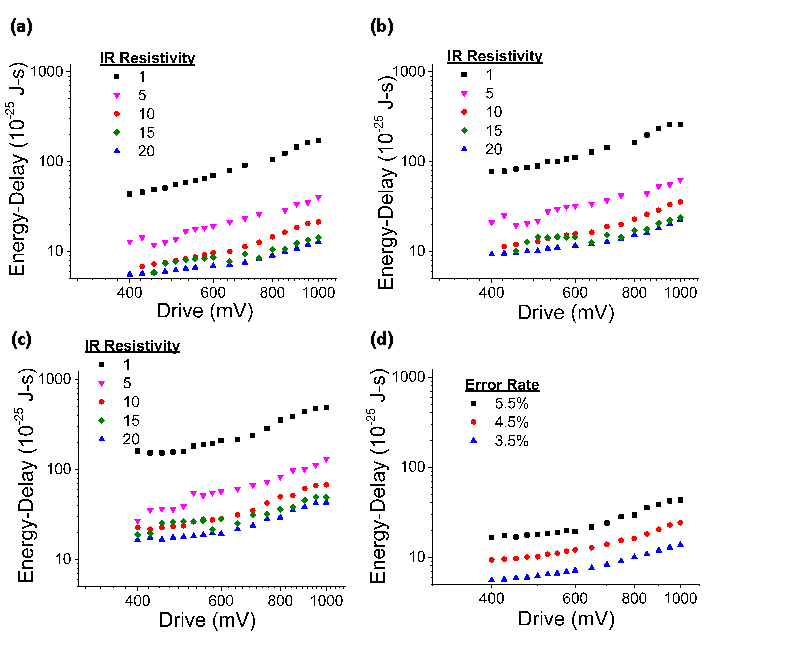}
\caption{ Average energy-delay product per cell vs. network drive potential with IR resistivity in m$\Omega \cdot$ cm and target error percentage as parameters. (a) 5.5$\%$ error. (b) 4.5$\%$ error. (c) 3.5$\%$ error. (d) All target error percentages at 20 m$\Omega$ $\cdot$ cm resistivity.}
\label{fig:ProductDriveData}
\end{figure}

The average energy-delay product per cell is plotted against the IR read path drive potential in Fig. \ref{fig:ProductDriveData}. Again, target error percentage and spin-orbit resistivity are used as parameters.

\begin{figure}
\centering
\includegraphics[scale=0.48]{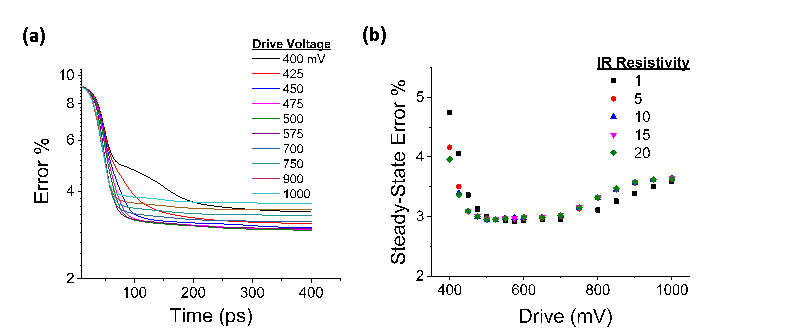}
\caption{ (a) Evolution of average system error over time with drive potential in mV as parameter. (b) Average final steady-state error vs drive potential with IR resistivity in m$\Omega \cdot$cm as parameter.}
\label{fig:TimeData}
\end{figure}

The evolution of the average error over time is shown directly in Fig. \ref{fig:TimeData} (a) for each drive potential level. The steady-state resting point of the error shows an intriguing trend. There are two regimes for low and high drive voltage. In the low drive regime the error reaches its steady-state level more quickly as drive increases. The steady-state error also decreases with increasing drive. At about 500 mV drive we transition to the second regime in which the steady-state error is not reached significantly faster with increasing drive. In addition the steady-state error actually increases with increasing drive. This is seen more clearly in Fig. \ref{fig:TimeData} (b). The final error initially falls, then rises more slowly with increasing drive. The optimum appears to be between 500-600 mV. The low drive regime is clearly a result of switching field limited operation. However it is less clear why the error should behave this way in the high drive regime. We believe this behavior to be due to the fact that a CNN cannot differentiate between correct and noisy pixels. As the drive becomes small, insufficient energy is available to switch any pixels. As the drive becomes large, pockets of noisy pixels surrounding `correct' pixels can more quickly influence their trapped neighbors. Recalling from Sec. \ref{sec:Results} A that the switching time decreases as the square of the drive voltage, it becomes more likely at high voltages that a fully trapped pixel will switch to agree with its neighbors before those partially surrounded neighbors are themselves corrected. The isolated neighborhood of noisy pixels may thus achieve sufficient size to maintain itself indefinitely, and will appear in the final corrected image. The range of 500-600 mV appears to be the `sweet spot' for preventing this behavior by slowing down pixel switching without lowering the input energy so far as to lose the ability to switch some pixels entirely.

\vspace{1in}

\section{Discussion}
\label{sec:Discussion}

\subsection{Thermal Stability}
The switching of very small magnetic vectors is a stochastic process heavily dependent on thermal fluctuations. Although stochastic CNNs are themselves a field of interest\cite{StochasticCNNs}, the IRMEN network is largely deterministic, as the external fields significantly outweigh thermal effects. This is necessary for the sake of nonvolatility. The FM dimensions were chosen in part to ensure this thermal stability. With a volume of $1536$ $nm^3$ and anisotropy $K = 6\cdot 10^5$ $erg/cm^3$ the magnets have an energy barrier of about $22kT$. With a reasonable attempt period of $\tau_0 = 1 ns$, the Neel-Arrhenius equation
\begin{gather}
\tau_N = \tau_0 \cdot e^{\frac{KV}{k_BT}}
\end{gather}
predicts that an IRMEN is unlikely to switch due to thermal fluctuations at room temperature on time-scales below seconds. This is far longer than the operational time scale of the network.

\subsection{Ferroelectric Properties}
Many recent Boolean devices involving magnetoelectric capacitors have assumed a significant degree of ferroelectric hysteresis in the magnetoelectric material\cite{COMET,MESO2,DWL}. This is beneficial for a digital logic circuit in which clock pulses can rotate the electric polarization and release it, leaving behind a remanent polarization which produces an effective magnetic field without the necessity of holding excess charge on the capacitor. However, in this CNN, there is no clock, as the neurons are meant to interact constantly. There is no need for a remanent polarization. Indeed, our model is simplest in the case of no ferroelectric hysteresis when a true linear relationship exists between applied voltage and polarization. A linear relationship is not critical to correct operation of an IRMEN network, however. With the assumption of a step-like magnetic transfer function as explained in Section \ref{sec:Simulation}, the ferroelectric hysteresis curve simply influences the critical input voltage at which the threshold polarization, and therefore critical field $H_{ME}$, is reached. We need only assume that the electric polarization is not saturated below this level. The relatively high saturation point of magnetoelectric materials makes this unlikely \cite{HighSatField,BFO2}.

\subsection{Scaling}
Here we will examine the effect of scaling on the relevant features of an IRMEN, in particular the IR and ME aspects. We see in (\ref{eq:RX}) that the efficiency of the IR conversion effect in the read path is proportional to $\frac{1}{F}$ where $F$ is the minimum feature size. This gain in IR efficiency from a reduced $F$ will ultimately enhance the magnetoelectric field by $\frac{1}{F^2}$ (\ref{eq:linearME}) and thus the neuron switching time. However the same equation also indicates that a reduction of the magnetoelectric capacitance due to reduced area will counterbalance the increased spin to charge efficiency to keep the effective ME field constant. As a result there is no change in the switching delay or the charging energy. This is in stark contrast to CMOS-based devices in which a reduced capacitor area leads to shorter delays because of the charging time. In this case, the bottleneck is the single-domain magnetic switching, limited by the area-independent Gillbert damping. We also consider the scaling of leakage power. Reducing the area of the read path increases the resistance resulting in less leakage power by a factor of $F^2$ within the neuron for constant drive voltage. On the other hand, the increase in IR efficiency mentioned above results in greater steady-state leakage from rail to rail by a factor of $\frac{1}{F}$ in the synapses. Finally, a purely negative result of scaling is a reduction of the FM thermal stability factor. These considerations lead to the conclusion that there is an optimal value of $F$ for the neuron based on the tradeoff of area, leakage and stability that should not be ventured past for a given material set. 

However, it is important to note that scaling need not be performed uniformly on the entire neuron structure. For instance, if neuron leakage and FM switching dominate the power and delay of a device, then scaling down the read path portions of the device while maintaining the ME capacitor stack at a constant area will yield a net advantage in both energy and speed. Furthermore, the area of an IRMEN cell is mostly taken up by the synapse FETs even in the simplest network with five synapses per cell. The benefits of scaling FETs are well-known. Thus significant reductions of synapse leakage power and overall cell area can be achieved simply by using the latest CMOS technology even with the neuron design unchanged.

\subsection{Architecture Comparisons}
Many non-Boolean spintronic computing schemes use a magnetic tunnel junction (MTJ) as the readout mechanism \cite{Naeemi1,Naeemi2,LIF,Ising}. One of the well-known failings of MTJs is their low resistance ratio which makes it difficult to differentiate the states. Encoding the output value as a resistance also means the output cannot swing between positive and negative values without using an additional layer of transduction, such as a voltage divider. These two MTJ shortcomings make the IR readout mechanism an interesting alternative. Taking advantage of efficient spin/charge transduction effects may allow for greater state differentiation. The IR readout also provides an output range which is symmetric about zero without requiring additional transduction layers since the projection of the magnetic orientation swings between $\pm1$. An additional advantage of the IRMEN CNN architecture over the Ising scheme in particular is the continuous nature of its computation. The network in \cite{Ising} is characterized by discrete write-relax-read pulses. In contrast, every IRMEN in a network is updated simultaneously, and each performs both its writing and reading operations at the same time. This parallelization of the neuron operations leads to an enhanced speed. 

\subsection{Energy-Delay Comparisons}
Perfect comparison of the energy efficiency between IRMEN CNNs and the aforementioned schemes is difficult as the architectures, functions and allowed error levels vary and the network performance depends on the size of the neuron. However a rough comparison indicates the IRMEN scheme is competitive and may even outperform the others. The best spintronic networks shown in\cite{Naeemi1} perform filtering with $10\%$ initial noise with an optimal energy-delay product in the range $3-8\cdot 10^{-24}$ J-s. This corresponds to a performance range of 0.5 ns and 6 fJ per cell per operation up to 2 ns and 4 fJ per cell per operation. The IRMEN network was simulated completing a filtering operation with the same initial noise to under 5.5$\%$ error in under 70 ps at a cost of under 10 fJ per cell corresponding to an energy delay product on the order of $7\cdot 10^{-25}$ J-s. Other spintronic networks were benchmarked using the associative memory function rather than filtering \cite{Naeemi2}. Based on the simulation of both functions on the same network by Naeemi \emph{et al}\cite{Naeemi1}, we estimate the enhanced complexity of this function results in an increase of approximately $3x$ for the required energy and time. Accordingly, we estimate the energy and delay of the IRMEN scheme for the associative memory operation to be approximately 30 fJ per neuron and 210 ps to complete. The spin-Hall and domain-wall networks in \cite{Naeemi2} perform at approximately 190 fJ and 8 ns, and 80 fJ and 10 ns respectively.

The spin-Hall based Ising cells use a total of 320 fJ for each 10 ns write-relax-read sequence\cite{Ising}. The ME-LIF neurons consume 18.5 fJ per read-reset operation \cite{LIF}. The authors report an additional 246 fJ consumed by each neuron during each training iteration, which we interpret as the write energy. For an upper bound on the cost and delay of a single read/write event we use the 10 fJ consumption of an IRMEN cell during an entire 70 ps network operation as given above. This demonstrates a significant improvement over the Ising and ME-LIF schemes, although since the latter is unique among the neurons mentioned here in that it represents a spiking neural network architecture which computes in the frequency domain, it may have advantages in other areas.

While direct comparison of the energy and delay between IRMEN and other spintronic neurons may not be precise for the reasons mentioned above, it is clear that at  optimal design points, the IRMEN performance can outperform other spintronic networks. In particular, even when the IRMEN cell consumes similar energy to its compatriots, its delay is significantly reduced. The IRMEN and other spintronic implementations both significantly outperform CMOS CNNs\cite{Naeemi1}. We note that this remarkable performance represents an upper bound, contingent somewhat upon the strength of the interlayer coupling strength $\zeta$ in Table \ref{tab:parameters}. Further experiments will determine the veracity of our assumption as to its value.  

\section{Conclusion}
\label{sec:Conclusion}
The impending end of CMOS scaling presents a unique challenge to researchers. The next generations of computational devices will depart from the past more starkly than ever before. Foreseeing this, researchers have redoubled work in both novel methods and computing schemes, including spintronics and neuromorphics respectively. We have designed and provided proof of concept for an inverse Rashba-Edelstein magnetoelectric device suited to neuromorphic computing with fast, low-power functions. We hope this work will provoke further investigation into materials with useful spin properties and provide an answer to the question of next-generation computing.

\end{document}